\documentclass{gen-p-l}
\usepackage{graphicx}
\usepackage{hyperref}

\theoremstyle{definition}

\theoremstyle{remark}

\numberwithin{equation}{section}

\begin{document}

\title[Impurities in Mott insulators]{Non-magnetic impurities as
probes of insulating and doped Mott insulators in two dimensions}

\author{Subir Sachdev}
\address{Department of Physics, Yale University, P.O. Box 208120,
New Haven CT 06520-8120, USA}
\email{subir.sachdev@yale.edu}
\urladdr{\href{http://pantheon.yale.edu/~subir}{http://pantheon.yale.edu/\~\/subir}}
\thanks{We thank H.~Alloul, J.~Bobroff, A.~Georges, and T.~Senthil for useful
discussions, and C.~Buragohain, A.~Polkovnikov, and Y.~Zhang for
collaborations on related work \cite{tolya,science,vojtaprl}.
Supported by NSF Grant DMR \#96--23181.}

\author{Matthias Vojta}
\address{Theoretische Physik III, Elektronische Korrelationen und
Magnetismus, Universit\"{a}t Augsburg, D-86135 Augsburg, Germany.}
\email{matthias.vojta@physik.uni-augsburg.de}

\date{September 10, 2000}


\begin{abstract}
We characterize paramagnetic Mott insulators by their response to
static, non-magnetic impurities. States with spinon deconfinement
(or spin-charge separation) are distinguished from those with
spinon confinement by distinct impurity susceptibilities and
finite-size spectra. We discuss the evolution of physical
properties upon doping to a $d$-wave superconductor, and argue
that a number of recent experiments favor spinon confinement in
the reference Mott insulating state.
\end{abstract}

\maketitle

\section{Introduction}
\label{intro} Soon after the discovery of high temperature
superconductivity, Anderson \cite{pwa} made the prescient
suggestion that the phenomenon is related to the physics of a
doped Mott insulator. The parent insulating compound, ${\rm La}_2
{\rm Cu O}_4$, is well described at low energies by the
excitations of a model of single orbitals on the vertices of the
square lattice at a density of one electron per site. The ground
state of this model is an insulator, and is known to have
antiferromagnetic long-range order. Nevertheless, Anderson argued
that the appropriate reference state was a paramagnetic Mott
insulator without antiferromagnetic long-range order, often
loosely referred to as a ``spin liquid''. In the intervening
years, much effort has been expended towards finding such spin
liquid states, and a number of definite candidates have emerged.
There are important qualitative distinctions between these
candidates, and we are especially interested in distinguishing
states which cannot be deformed adiabatically into each other,
and must be separated by a quantum phase transition. In
particular, a key propery is whether the state allows deconfined
$S=1/2$ spinon excitations or not. If it does, then these
neutral, $S=1/2$ excitations imply that ``spin-charge
separation'' has occurred.

In this paper, we argue that the response of the ground state to
non-magnetic impurities is a sensitive probe of spinon
confinement, and so is a central distinguishing characteristic of
the various paramagnetic Mott insulators. Experimentally, such
non-magnetic impurities can be easily created by substituting Zn
or Li on the Cu sites, and a large number of such experiments
have been carried out on both the insulating and superconducting
compounds. We will describe the results of some of these
experiments here, and argue that they offer evidence that the
appropriate reference Mott insulating state to the high
temperature superconductors is one in which the spinons are
confined; our point of view is therefore contrary to \cite{pwa2}.
We will also mention theoretical work \cite{rs1,book} which argues
that such confined states break translational symmetry by the
appearance of spin-Peierls order (which can also be viewed as a
``bond-centered stripe'').

The spin excitations of a Mott insulator are usually well
described by the Heisenberg Hamiltonian
\begin{equation}
H = \sum_{i<j} J_{ij} {\bf S}_i \cdot {\bf S}_j + \ldots
\label{e1}
\end{equation}
where ${\bf S}_i$ are SU(2) operators with spin $S=1/2$ on the
sites, $i$, of some lattice, and the ellipsis represent possible
additional multiple spin exchange terms. We will describe cases
in which $H$ has a paramagnetic ground state with confined
spinons in Section~\ref{confine} and also discuss their response
to non-magnetic impurities. The corresponding discussion for
models with deconfined spinons is in Section~\ref{deconfine}.
Finally Section~\ref{dwave} uses the insights gained to review
recent experimental and theoretical work on non-magnetic
impurities in $d$-wave superconductors.

\section{Confined spinons}
\label{confine}

\begin{figure}
\centerline{\includegraphics[width=4in]{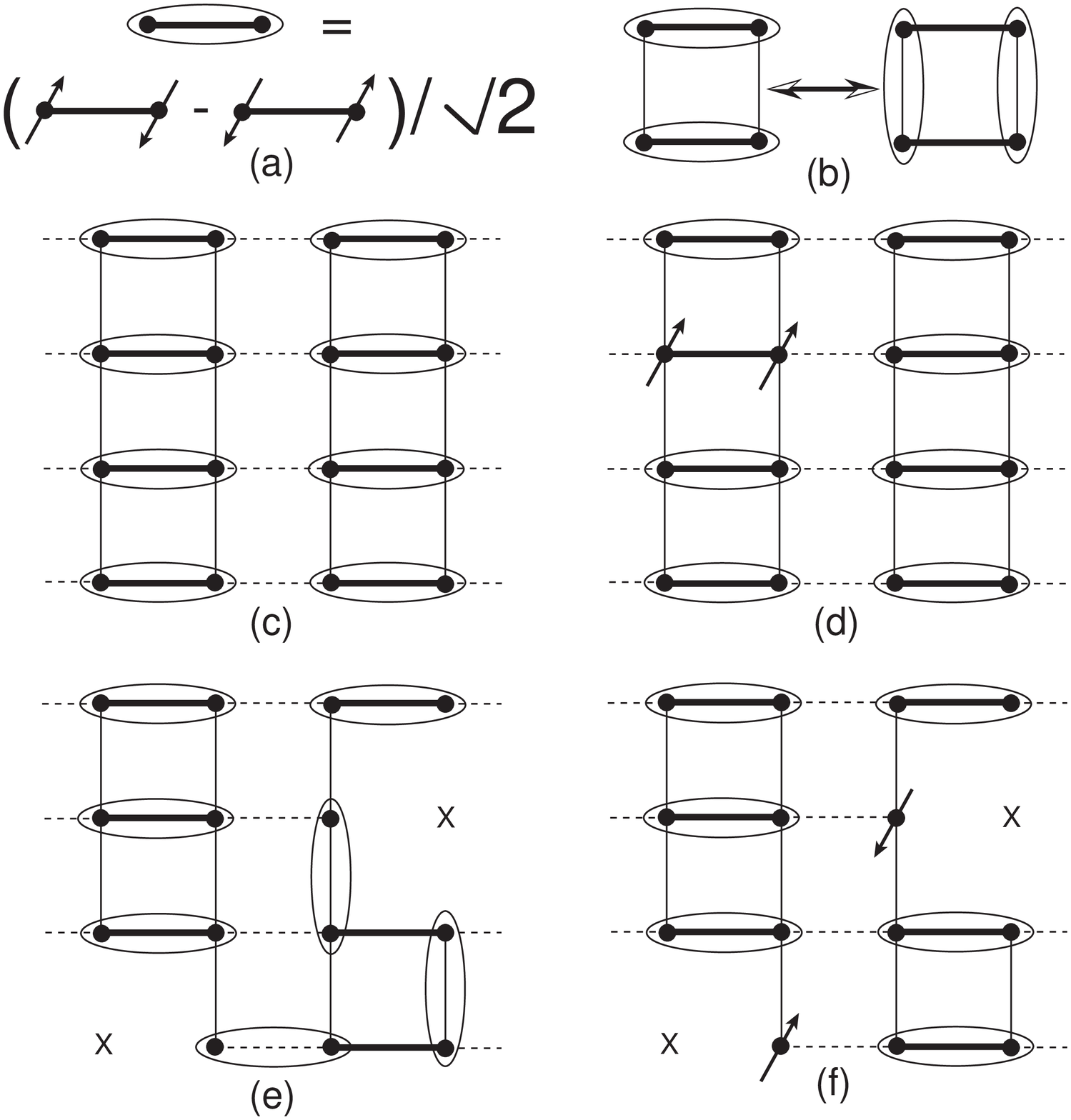}} \caption{(a)
Definition of an ellipse as the representation of a singlet
bond between a pair of sites. (b) Two different singlet pairings
(``dimer packings'') of pairs of spins around a plaquette. The
resonance between these will be largest when the exchange
constants represented by the think and think straight lines are
equal. (c) Spin ladder realization of $H$. The thick, thin, and
dashed straight lines represent $J_{ij}>0$ couplings of differing
values. A snapshot of the ground state is shown for the case
where the thick lines have the largest $J_{ij}$. (d) $S=1$
particle excitation of (\protect\ref{disp1}) represented by the
absence of a singlet bond between a pair of spins; the two free
spins propagate throughout the lattice but remain confined to
each other. (e,f) Two candidate ground state configurations in the
presence of two non-magnetic impurities (represented by the X's).
For a confining paramagnet, the configuration in (f), with two
free moments near the impurities, is always preferred once the
impurities are sufficiently far apart.} \label{fig1}
\end{figure}

Consider the spin-ladder realization of $H$, as defined in
Fig~\ref{fig1}c.
For the case where the thick lines have $J_{ij}$ much larger than
all other exchanges, the state in Fig~\ref{fig1}c is a good
approximation to the ground state. As the values of $J_{ij}$ on
the three sets of links become equal to each other, there is
increasing resonance in the ground state between other singlet
pairings between the sites, until at a critical point there is an
onset of magnetic long-range order. However, the nature of the
paramagnetic state remains essentially unchanged all the way up
to the critical point. In particular, the lowest-lying excitation
is a $S=1$ particle  which is shown schematically in
Fig~\ref{fig1}d: the columnar pairing of the spin singlets in the
environment ensures that the two spin-1/2's at the ends of the
broken bond cannot move apart from each other, {\em i.e.}, the
spinons are confined. If we represent the dispersion of the
stable $S=1$ particle by
\begin{equation}
\varepsilon_k = \Delta + \frac{c_x^2 k_x^2}{2 \Delta} +
\frac{c_y^2 k_y^2}{2 \Delta} \label{disp1}
\end{equation}
($k = (k_x, k_y)$ is the momentum, $\Delta$ is the gap to spin
excitations, and $c_{x,y}$ are velocities), then the response to
an external magnetic field is determined entirely by the
thermally excited density of such particles. A simple calculation
then shows that the susceptiblity, $\chi_u$, of a sample of area
$\mathcal{A}$ is given by \cite{book}
\begin{equation}
\chi_u = \frac{\mathcal{A} \Delta}{\pi c_x c_y} e^{-\Delta/T},
\label{chiu}
\end{equation}
where $T$ is the absolute temperature, $\hbar=k_B=1$, and we have
absorbed factors of the electron magnetic moment into the
definition of the suscpetibility.

It is important to keep in mind that the above picture of the
confined paramagnet holds not only for anisotropic spin ladders
just mentioned, but also for isotropic models in which the
Hamiltonian has the full four-fold rotational symmetry of the
square lattice about every site \cite{rs1,sushkov}. In this case,
the model with only nearest neighbor interactions is known to
have magnetic long-range order, and so frustrating second
neighbor interactions are necessary to access the paramagnetic
state. It has been argued that this state {\em spontaneously\/}
breaks the square lattice rotational symmetry so that the pattern
of singlet bonds in one of the four equivalent ground states has
the same symmetry as the configuration in Fig~\ref{fig1}c
\cite{however}. This broken symmetry can be understood in the
framework of a ``quantum dimer'' model \cite{dimer} of resonating
nearest-neighbor singlet bonds: among all the dimer packings, the
columnar pattern of dimers has the maximum number of states which
can resonate with it by flipping a pair of singlet bonds as shown
in Fig~\ref{fig1}b. More technically, the quantum dimer model is
dual to a compact U(1) gauge theory \cite{rs1,frad,curreac}, and
the confining property of this theory implies that translational
symmetry is broken. The non-zero spin excitations of such a
paramagnet continue to be described by the $S=1$ particle in
(\ref{disp1}).

Let us now add a non-magnetic impurity. Actually, it is
convenient to always deal with systems with an even number of
spins, and so we will add two impurities and eventually move them
infinitely far apart from each other. These two impurities are
represented by the two X's in Figs~\ref{fig1}e,f. If we now
attempt to construct the ground state of the system with
impurities, two distinct possibilities exist. First, we can
remain within the subspace of short-range singlet bonds ({\em
i.e.}, the Hilbert space of the quantum dimer model), and this is
indicated in Fig~\ref{fig1}e. The key property of such a state is
that there is a string of `defect' bonds connecting the two
impurities which is out of registry with the global columnar
order; so as the impurities move apart from each other, there is
an energy cost which grows linearly with the separation between
the impurities. This linear energy cost can also be understood
within the compact U(1) gauge theory representation of the dimer
model, in which the impurities at X appear as static electric
charges which are linearly confined by the gauge force. So
ultimately, it will always pay for the system to break a singlet
bond, and produce two nearly free moments (`spinons') confined
around each impurity \cite{fink,rs1,sy,sigrist,imada,fabrizio},
as shown in Fig~\ref{fig1}f. There is a weak effective
interaction, $J_{\rm eff}$, between these moments which is
mediated by virtual excitations of the intervening singlet bonds,
and is therefore exponentially small in the separation, $R$,
between the impurities. We expect $|J_{\rm eff}| \sim \Delta
e^{-R \Delta/c}$ where $c$ is of order the geometric mean of
$c_x$, $c_y$. The sign of $J_{\rm eff}$ will be (anti-)
ferromagnetic if the impurities are on the same sublattice
(opposite sublattices).

The presence of these two weakly interacting moments has strong
and distinctive signals in the spectrum and thermodynamics of the
model. Upon exact diagonalization of a finite-size system of two
impurities, one should find a very low-lying $S=1$ state, well
below the bulk spin gap, which approaches the ground state
exponentially fast as the separation between the impurities is
increased. For $J_{\rm eff}<0$, the $S=1$ state will eventually
become the global ground state. In the presence of a uniform
external magnetic field, the free moments contribute a large
susceptibility which is easily detectable experimentally; in
addition to the bulk contribution in (\ref{chiu}), we have the
impurity contribution
\begin{equation}
\chi_{\rm imp} = \frac{2}{T}\frac{ e^{-J_{\rm eff}/T}}{1+3
e^{-J_{\rm eff}/T}}. \label{chiimp}
\end{equation}
This has a Curie divergence of two free moments, $\sim 1/2T$, for
$T>|J_{\rm eff}|$ which excludes only an exponentially small
low $T$ regime as $|J_{\rm eff}|$ is so small.

\section{Deconfined spinons}
\label{deconfine}

\begin{figure}
\centerline{\includegraphics[width=3in]{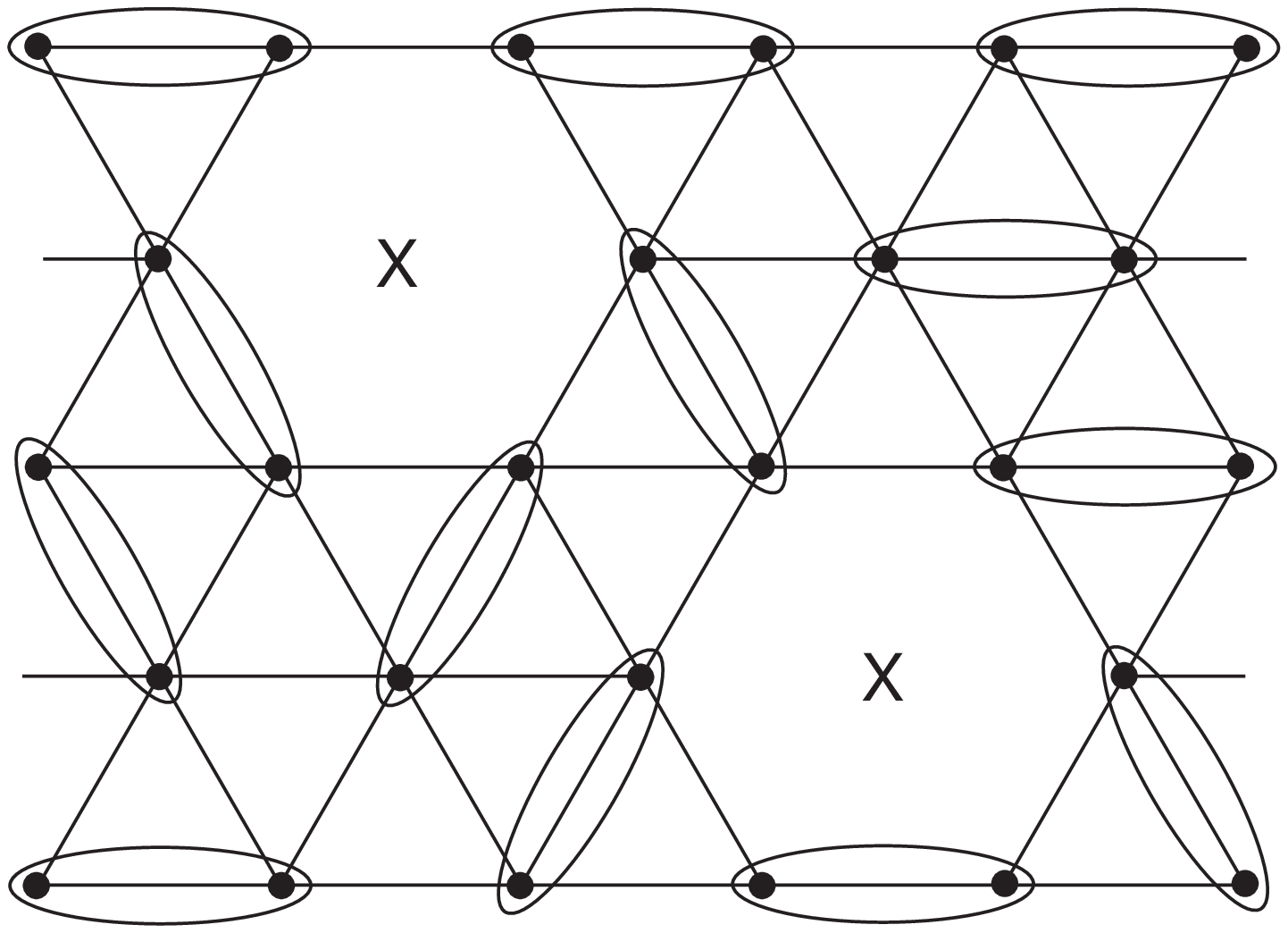}}
\caption{Snapshot of the wavefunction on the triangular lattice
in the presence of two impurities. The two impurities are not
connect by a line of ``defect'' bonds and are free to move
infinitely apart from each other.} \label{fig2}
\end{figure}

We now look for paramagnetic Mott insulators in which non-magnetic
impurities do not bind local moments in their vicinity. Clearly,
the key effect responsible for confinement in
Section~\ref{confine} was the rigidity of the wavefunction in the
space of different singlet bond pairings, produced by the energy
gained in the resonance of Fig~\ref{fig1}b. We can anticipate that
stronger fluctuations will appear in this singlet pairing space by
allowing frustrating exchange interactions in $H$, {\em e.g.},
diagonal or second-neighbor bonds on the square lattice will
permit resonance around a much bigger class of loops, some of
which overlap with each other. More technically, we need the
quantum dimer model to be dual to a deconfined gauge theory: the
electric charges induced by the non-magnetic impurities will then
not be confined to each other, and it will be possible to move
the impurities apart with negligible energy cost once they are
well separated. In such a situation, even in the presence of the
impurities, the ground state will remain within the subspace of
short-range singlet bonds, and no nearly-free spinons will be
generated. A schematic of such a state with impurities is
indicated in Fig~\ref{fig2} for the triangular lattice.

Such paramagnetic Mott insulating states with strong fluctuations
in the space of singlet pairings were discussed by a number of
investigators under the general umbrella of ``resonating valence
bond'' (RVB) states \cite{faz,pwa,sutherland,dimer,bulbul}.
However, a clear distinction between deconfined RVB wavefunctions
and the confined states in Section~\ref{confine} was not made. A
specific condition for deconfinement was spelt out in
Refs.~\cite{rs2,rs2d,wen}: one needed condensation of a charge
$\pm 2$ Higgs field, realized by a dimer on a link with a
frustrating interaction, to move the compact U(1) gauge theory
into a deconfined phase (the quantum transition between the
confined and deconfined states is described by a $Z_2$ gauge
theory \cite{js2,senthil,curreac}). Examples of such deconfined
phases where presented in large $N$ computations on frustrated
antiferromagnets on the square \cite{rs2,rs2d}, triangular, and
kagome lattices \cite{kagome}. Theses phases also have stable,
spin-singlet, $Z_2$ vortex excitations \cite{rs2,book}
(christened `visons' in recent work \cite{senthil}), and it is
possible that the non-magnetic impurities bind visons in their
vicinity\cite{bulbul}; even so, the impurities will be able to
move apart as there is no long-range force between two visons.
Formation of local moments near the impurities will require the
breaking of singlet bonds, and this is suppressed by the presence
of the spin gap. Of course, even in the absence of confinement,
we cannot completely rule out the possibility that there is an
impurity-spinon bound state lower in energy than an isolated
impurity and a impurity-vison bound state, but we consider it
unlikely in a situation where short-range RVB states are strongly
preferred.

Recently, a couple of plausible candidates for deconfined states
have emerged in numerical studies of specific models on the
triangular lattice: Misguich, Lhuillier and
collaborators \cite{clair} examined an extention of $H$ with
ring-exchange interactions, while Moessner and
Sondhi \cite{moessner} argued for a deconfined phase for the
quantum dimer model.

We close this section by contrasting the spectral and
thermodynamic properties of confined states (discussed in
Section~\ref{confine}) with the corresponding properties of
deconfined states. All specific models for deconfined states
discussed so far have been found to have a {\em pair} of $S=1/2$
spinon excitations \cite{rs2,book}; the two spinons have minima
for their dispersion at different points in the Brillouin zone,
and the dispersion is described by (\ref{disp1}) in its vicinity.
As in (\ref{chiu}) we can then compute the susceptibility to a
uniform magnetic field
\begin{equation}
\chi_u = \frac{\mathcal{A} \Delta}{2 \pi c_x c_y} e^{-\Delta/T},
\label{chiu1}
\end{equation}
which differs from (\ref{chiu1}) by a factor of 1/2 (also
$\Delta$ is now the gap to $S=1/2$, rather than $S=1$
excitations). A more striking difference appears when we consider
the response to two non-magnetic impurities: no local moments
form, and so there is no low-lying $S=1$ state exponentially
close to the ground state. Consequently the change in the
susceptibility is not large: it is expected to be of order
(\ref{chiu1}) with $\mathcal{A}$ replaced by the area of two unit
cells. The absence of low-lying $S=1$ states should also serve as
a sensitive diagnostic of deconfinement in numerical studies: it
should be interesting to extend the numerical results in
\cite{clair} in this direction.

\section{$d$-wave superconductors}
\label{dwave}

The Bogoliubov quasiparticles of a superconductor have
essentially the same quantum numbers as deconfined
spinons \cite{kivelson}, and so there is no fundamental reason why
a non-magnetic impurity in a good $d$-wave superconductor must
necessarily bind a $S=1/2$ moment in its vicinity. However, when
one considers incrementally doping a paramagnetic Mott insulator,
two distinct possibilities arise.

\noindent (A) For the deconfined states of
Section~\ref{deconfine}, both the weak and strong doping limits
do not bind moments near the impurity: consequently, we  expect a
smooth evolution of physical properties with doping, with only a
weak magnetic response associated with the impurity.

\noindent (B) A quite different picture emerges upon doping the
confined states of Section~\ref{confine}. The undoped limit has a
$S=1/2$ near each impurity, while the strongly doped limit does
not: we expect that the free $T=0$ moment will survive in the
superconducting state for a finite range of doping ({\em i.e.}, an
isolated moment will exhibit a divergent Curie susceptibility
$\sim 1/4T$ even in the superconducting state), and a quantum
critical point separates the weak and strong doping limits. On
the strong doping side of this quantum critical point, the moment
is Kondo screened as $T \rightarrow 0$, and such a regime is
continuously connected to a regime, at higher doping, where the
moment does not even form at intermediate $T$. This quantum phase
transition is described by a Kondo-like Hamiltonian of a $S=1/2$
moment exchange-coupled to the gapless Bogoliubov quasiparticles.
This model has been much studied \cite{kondo,tolya,zhu} in recent
years, and it is well established that such a transition does
exist in models without particle-hole symmetry; however, no
fundamental understanding of the nature of the critical
properties of the transition has yet been achieved.

We wish to argue here that a number of recent experiments on the
high temperature superconductors suggest that possibility (B) is
the correct  one, {\em i.e.}, the appropriate reference Mott
insulating state in the one with spinon confinement, and that
non-magnetic impurities in the under-doped superconductor always
bind a $S=1/2$ moment at $T=0$. The most direct evidence comes
from recent NMR measurements of Alloul and collaborators
\cite{alloulnew} who have measured a Curie-like susceptibility of
Li moments in the {\em superconducting} state of underdoped YBCO;
a number of earlier experiments \cite{fink,alloulold} have also
seen moments above the superconducting critical temperature in the
normal state. Indirect, but strong, evidence comes from our
studies of the influence of the non-magnetic impurities on other
excitations of the superconductor:\\ ({\em i\/}) the resonant
$S=1$ spin collective mode broadens dramatically upon doping with
a very small concentration of Zn impurities \cite{keimer}, and we
have argued \cite{science} that this is naturally explained by the
unpaired moments near the impurity. Indeed, the very existence of
the spin resonance mode is evidence for confinement in the
reference insulating state, as the resonance may be viewed
\cite{lt} as the continuation of the $S=1$ particle discussed in
Section~\ref{confine} and Fig~\ref{fig1}d.\\
({\em ii\/}) STM experiments measuring the quasiparticle
tunneling current near Zn impurities in BSCCO have an unusual
dependence on bias voltage and spatial location \cite{seamus}, and
this is reproduced \cite{tolya} in the Kondo-like models mentioned
above.

Finally, we mention the issue of translational symmetry breaking.
We have argued that the translational symmetry breaking must be
present in the confined insulator, and it is natural to expect
that this will survive in at least the lightly-doped
superconducting state\cite{rs2,rs2d,vojtaprl}, and a state with
co-existing stripe and superconducting order was discussed early
on in \cite{rs2d}. Indeed, it is tempting to relate the stripes in
Fig~\ref{fig1}c to the stripes experimentally observed in the
superconductor \cite{vojtaprl,curreac}. However, at least at very
low doping, it seems clear that the magnetic long-range order
present in the actual undoped insulator plays an important role
in the microstructure of the stripes. But at somewhat larger
doping, once the magnetic order disappears, we think it is
plausible that the local stripe correlations are bond-centered
and look similar to those in Fig~\ref{fig1}c. Indeed, a recent
photoemission experiment \cite{shen} has suggested bond-centering
of stripes in this regime.

We reiterate that while local moment formation near non-magnetic
impurities is intimately linked with confinement and
translational symmetry breaking in the insulating,
charge-incompressible state, this is no longer expected to be the
case in the superconductor. The latter is compressible and local
moments may be present even in a phase without translational
symmetry breaking. The main argument for moment formation in the
superconductor is based on continuity from the undoped insulator;
the quantum critical point at which the moment is quenched with
increasing doping can be distinct from the critical point at which
bulk translational and rotational symmetries are restored (and
also from the point where the ground state becomes a normal metal
at large overdoping).

\bibliographystyle{amsalpha}

\begin{thebibliography}{99}
\bibitem{however} An alternative plaquette order is also possible
(see~\protect\cite{book} and~\protect\cite{sushkov} and
references therein), but its response to magnetic impurities is
very similar to that discussed in Section~\protect\ref{confine}.
\end{thebibliography}

\end{document}